# Ligand co-deposition in focused electron beam induced nanoprinting: a predictive composition model

*Jakub Jurczyk[1,2,3]\*, Leo Brockhuis[1], Amalio Fernández-Pacheco[1] and Ivo Utke[2]\**

J. Jurczyk, A. Fernández-Pacheco

Institute of Applied Physics, TU Wien, Vienna, Austria

jakub.jurczyk@tuwien.ac.at

J. Jurczyk, L. Brockhuis, I. Utke

Laboratory for Mechanics of Materials and Nanostructures, EMPA Swiss Federal

Laboratories for Materials Science and Technology, Thun, Switzerland

ivo.utke@empa.ch

J. Jurczyk

Faculty of Physics and Applied Computer Science, AGH University of Cracow, Kraków,

Poland

Funding: Horizon 2020 Program, Contract No. 101001290, 3DNANOMAG, EU Horizon 2020 Marie Curie-Sklodowska Innovative Training Network "ELENA", grant agreement No 722149; Swiss National Science Foundation, COST-SNSF project IZCOZ0_205450.

**Abstract**

Recent advances in nanotechnology have created the need to manufacture three-dimensional nanostructures with controlled material composition. Focused Electron Beam Induced Deposition (FEBID) is a nanoprinting technique offering highest spatial resolution combined with the ability to directly 3D-print almost any shape. It relies on local electron-induced dissociation of metal-ligand organometallic molecules adsorbed onto a substrate.

So far FEBID continuum modelling involved the surface kinetics of precursor molecules during electron irradiation and succeeded in the prediction of nanoprint shape and growth rate and forms nowadays the basis of software for 3D nano-printing of nanostructures.

Here, we expand the model to the surface kinetics of detached ligands. Involving their dissociation and desorption behavior allows to predict trends in the metallic composition of the nanoprinted material and to define desirable nanoprint process windows as function of electron exposure time and flux.

We present the theoretical foundations of the model, validate it experimentally for chromium and silver precursors, compare calculated values with literature data for various precursors, and discuss its potential to design new experiments. This contribution enhances our understanding of FEBID dynamics and provides a versatile framework for predictive FEBID material nano-printing.

# 1. Introduction

Focused electron beam induced deposition (FEBID) is a method for direct three-dimensional nanoprinting, allowing for the deposition of structures of nearly any shape without using complex, expensive masks [1]. The lateral sizes of the structures which can be produced using this technique are comparable with the size of the electron beam, reaching down to single nanometers [2]. FEBID is usually used for manufacturing metallic structures for applications in photonics and plasmonics [3, 4], magnetism [5, 6] or probing and sensing devices [7-10]. On an industrial scale, FEBID is applied to correct defects in photolithography masks [11].

The principles of the method are simple: gaseous, usually organometallic precursor molecules are continuously introduced into the vacuum chamber of a scanning electron microscope (SEM) and directed to the substrate. The physisorbed molecules are locally dissociated by the focused electron beam. Non-volatile fragments stay on the substrate creating a deposit. Volatile moieties desorb and are pumped out of the chamber. In an ideal case, only metal atoms stay on the surface and ligands are detached and evacuated [12]. However, one of the main constraints of the technique is the purity of the resulting deposits, with metallic compositions close to 95 – 100 at.% achieved only for few substances: Co with $Co(CO)_8$ [13, 14], Au with $(PF_3)AuCl$ [15] and $(CO)AuCl$ [16], Fe with $Fe(CO)_5$ [17] and Si[18]. In most cases, typical metallic content remains, however, between 10 and 20 at.% [19-21].

There are two main mechanisms responsible for a resulting low metal content: co-deposition of detached ligands and incomplete dissociation of precursor molecules [12]. Each of these two unwanted effects are caused by different process parameters. For example, if the average desorption time of detached ligands is long, compared to the pixel exposure time, these can be further dissociated by the electron beam, creating non-volatile fragments that are incorporated into the growing material. In another case, if the electron flux is too small, part of the ligand may stay attached to the metal core and be buried underneath fresh incoming precursor molecules. In general, mutual relations between different physical and technical parameters of the nanoprint process influence composition, size and shape of the material. An example of such relations is the influence of pixel dwell time (electron exposure time) and pixel pitch distance on growth rate and shape of deposited structure. These kind of mutual relations are crucial for preparing deposits with complex shapes and stand as fundaments for 3D pattern creation software like 3BID [22], the one developed by Huth et al. [23, 24], and f3ast [25].

Until now, FEBID modelling has been focused on single reaction product (or single step) frameworks, with the objective of reproducing deposition rate and shape of deposits rather than their composition. Multi-reaction product modelling has been performed only in a few cases: For example, it has been used for focused electron beam induced etching of Si with $NF_3$, where electron induced reactions involving the addition of fluorine to silicon to form volatile $SiF_4$, and the removal of fluorine from $SiF_n$ to form non-volatile Si, were involved [26, 27]. For deposition using FEBID, multi-reaction product modeling has been so far restricted to scenarios where two different precursors impinge on the surface and yield a certain deposit composition [28]. In this work, we develop an analytical two-reaction FEBID model with one single precursor, which is by far the most common scenario in FEBID. This model is an extension of the FEBID continuum model, aiming to properly describe co-deposition of detached ligands alongside the metal atoms. The model is compared to experimental data which cannot be explained by current models. The work presented here is significant for the development of functional 3D nanostructures, as it enables modeling not only of the growth rate but also of the composition of the deposited material. Precise control over both the shape and composition of the deposits will advance FEBID further as a leading method for 3D printing of functional nanostructures.

## 2. Single-species FEBID continuum model

There have been so far several ways to mathematically model FEBID processes [22, 26, 28-30]. By now, the main one is the FEBID continuum model [26]. This model is usually (although not necessarily [29]) based on a Langmuir assumption to describe an adsorption term (adsorption of maximum one monolayer of molecules) and describes the deposition of the surface adsorbed molecule $(ML)$ by the electron beam $e^-$ with a certain yield $Y_{ML}$:

$$(ML) + e^- \xrightarrow{Y_{ML}} M\downarrow + L\uparrow \qquad \text{(reaction 1)}$$

So far, the model has worked under the assumption of an instant desorption of the detached part L (usually an organic ligand or part of it) as symbolized by the upward pointing arrow; it also assumes perpetual deposition of the cleaved part M (usually containing the metal atom M) as symbolized by the downward pointing arrow. The deposition yield will depend on the surface concentration of the two reaction partners. The concentration of ML on the surface is obtained by the adsorption rate equation which considers four kinetic processes: adsorption, thermal desorption, diffusion and electron-induced dissociation. The surface concentration of molecules $n$ [m$^{-2}$] is described by the adsorption rate equation

$$\frac{\partial n}{\partial t} = J\left(1 - \frac{n}{n_0}\right) - \frac{n}{\tau} + D\left(\frac{\partial^2 n}{\partial x^2} + \frac{\partial^2 n}{\partial y^2}\right) - \sigma f n \qquad \text{(eqn. 1a)}$$

where $J$ [m$^{-2}$s$^{-1}$] is the gas flux of impinging precursor molecules, $n_0$ [m$^{-2}$] is the surface concentration of molecules for full monolayer coverage -and usually defined as the inverse of the adsorbed molecule area, $\tau$ [s] is the average desorption time, $D$ [m$^2$s$^{-1}$] is the surface diffusion coefficient, $\sigma$ [m$^2$] is the cross section of electron-induced dissociation, and $f$ [m$^{-2}$s$^{-1}$] is the electron flux. This model has been successfully applied to derive different deposition regimes and scaling laws, connecting size and shape of deposits with the parameters used for deposition [31]. To generalize the mathematical description from materials and FEBID parameters, characteristic rates can be defined [29], having inverse time as unit, for gas impingement $v_{gas} = J/n_0$, thermal desorption $v_{des} = 1/\tau$, and electron induced dissociation $v_{dis} = \sigma f$, which transforms equation 1a into

$$\frac{\partial \theta}{\partial t} = v_{gas}(1 - \theta) - v_{des}\theta - v_{dis}\theta + D\nabla^2\theta \qquad \text{(eqn.1b)}$$

where $\theta = n/n_0$ is the surface coverage. Neglecting the surface diffusion term allows to derive analytical solutions for the coverage and deposition yield (deposited atoms per impinging electron) [26, 29].

The coverage as function of electron beam exposure (dwell) time $t$ is obtained as

$\theta(t) = (\theta_0 - \theta_\infty)e^{-v_\Sigma \cdot t} + \theta_\infty$ (eqn. 2)

with the steady state coverage $\theta_\infty = \theta_{t\to\infty} = \frac{v_{gas}}{v_{gas}+v_{des}+v_{dis}}$, the initial coverage $\theta_0 = \theta_{t\to 0} = \frac{v_{gas}}{v_{gas}+v_{des}}$, and the sum rate $v_\Sigma = v_{gas} + v_{des} + v_{dis}$.

The unitless time averaged reaction (deposition) yield $Y$ is defined as

$Y = \frac{Y_{max}}{t_d} \int_0^{t_d} \theta(t) dt$ (eqn. 3)

with $Y_{max} = \sigma n_0$ and $t_d$ the dwell time. Inserting eqn. 2 gives for the temporal evolution of the deposition yield [26, 29, 31]:

$Y = Y_{max} \left\{ \frac{\theta_0 - \theta_\infty}{v_\Sigma \cdot t} [1 - e^{-v_\Sigma \cdot t}] + \theta_\infty \right\}$ (eqn. 4)

with $Y_{max} = \sigma n_0$, and the values for steady state

$Y(t \to \infty) = Y_\infty = Y_{max}\theta_\infty$ (eqn 4a)

and initial state

$Y(t \to 0) = Y_0 = Y_{max}\theta_0$. (eqn 4b)

For completeness we mention the deposition yield can be transformed to the experimentally accessible FEBID rate $R$ (in units of dimension over time) via the relation

$R = YVf$ (eqn. 5)

where $V$ represents the atomistic volume of the deposited material composition and $f$ the electron flux.

**3. Ligand co-deposition model**

The conceptual idea of the FEBID model presented here is shown in Figure 1. Precursor molecules (ML) adsorb on the substrate where they desorb or dissociate by impinging electrons.

After dissociation, precursor molecules are separated into non-volatile metal atoms M (or a metal containing part) and the detached ligand molecules L (or a fragment part of the original ligand). The detached ligands desorb or are dissociated by further impinging electrons. Upon their dissociation they result in a co-deposited non-volatile product $\Lambda$ and a volatile product $V$ that is treated as instantaneously desorbing and being pumped out. Surface diffusion is not included in the model to facilitate analytical solution of the equations.

The model presented here was inspired by observations on FEBID with silver [32, 33] and copper carboxylates[34]. The resulting granular deposits contained large (> 10 nm) metal grains and the metal content in the deposits varied inversely with the intensity of the electron flux, suggesting that ligand desorption is not instantaneous. Both observations suggest that ligands are detached from the metal atoms and can be co-deposited alongside them. A more detailed comparison of the ligand co-deposition model to experimental data will be given in section 5.

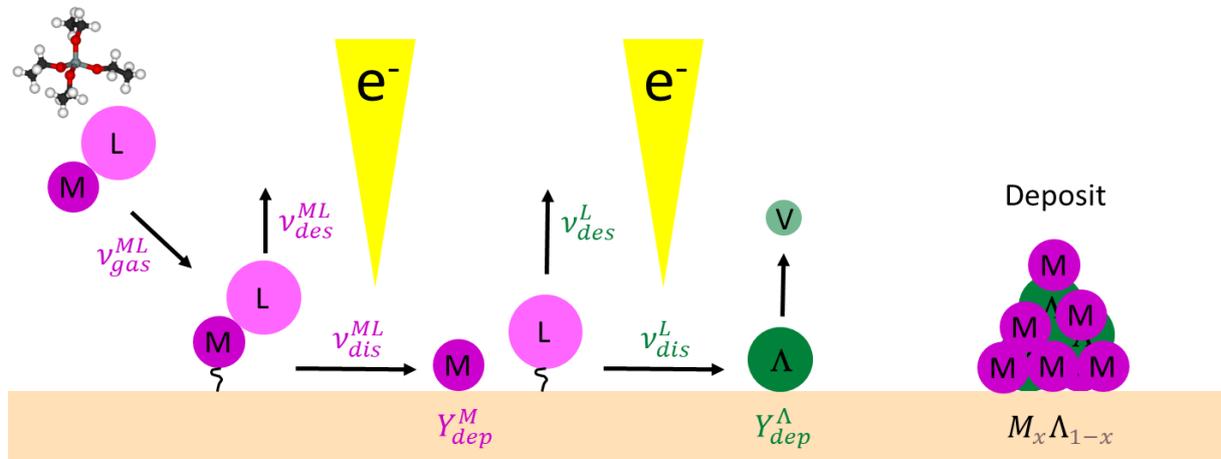

**Figure 1.** Schematic representation of the multi-step ligand co-deposition model. The precursor molecule is treated as metal atom (M) having a ligand (L) and reaches the substrate surface at a rate $\nu_{gas}^{ML}$ where it physisorbs. It can then desorb at a rate $\nu_{des}^{ML}$. Upon electron beam irradiation, it dissociates at a rate $\nu_{dis}^{ML}$, leaving behind the metal fragment. The efficiency of this process is described by the deposition yield $Y_{dep}^{M}$. The detached ligand (L) desorbs at a rate $\nu_{des}^{L}$ or undergoes further electron dissociation at a rate $\nu_{dis}^{L}$ into a non-volatile ligand residue $\Lambda$ and a volatile part V leaving instantaneously. The efficiency of ligand deposition is described by $Y_{dep}^{L}$. The final deposit consists of both metal atoms and deposited ligand residues, with a composition given by $M_x\Lambda_{1-x}$.

The model considers two electron-induced reactions with their respective yields $Y$:

$$(ML) + e^- \xrightarrow{Y_{ML}} M\downarrow + (L) \qquad \text{(reaction 2a)}$$

$$(L) + e^- \xrightarrow{Y_L} \Lambda\downarrow + V\uparrow \qquad \text{(reaction 2b)}$$

Reaction 2a describes the dissociation of the adsorbed metal-ligand containing molecule $(ML)$ into the deposited metal atom $M\downarrow$ and the adsorbed detached ligand $(L)$. Reaction 2b describes the dissociation of the adsorbed detached ligand $(L)$ into the deposited ligand residue $\Lambda\downarrow$ and the volatile part $\uparrow$. Desorption of $V$ is considered instantly. The formalism also includes incomplete dissociation of ligands from the metal atom by assigning a certain composition to the non-volatile and adsorbed detached products in reaction 2a. A set of two differential adsorption rate equations describes the coverages of precursor molecules $\theta^{ML}$ and ligands $\theta^L$ for reactions 2a&b. Eqn. 6a describes the temporal evolution of intact metal-ligand precursor molecules, and eqn. 6b considers the one of the detached ligand molecules on the substrate. The index $ML$ indicates the intact precursor molecule and index $L$ the adsorbed detached ligand:

$$\frac{\partial \theta^{ML}}{\partial t} = v_{gas}^{ML}(1 - \theta^{ML} - \theta^L) - v_{des}^{ML}\theta^{ML} - v_{dis}^{ML}\theta^{ML} \quad (6a)$$

$$\frac{\partial \theta^L}{\partial t} = \alpha v_{dis}^{ML}\theta^{ML} - v_{des}^L\theta^L - v_{dis}^L\theta^L \quad (6b)$$

where coverages of precursor molecules and detached ligands are defined as $\theta^{ML} = n^{ML}/n_0^{ML}$ and $\theta^L = n^L/n_0^L$ with $n_0^{ML}$ and $n_0^L$ being the complete monolayer surface concentrations of precursor molecules and detached ligands, respectively, and being inversely proportional to their respective area. The ratio $\alpha = n_0^{ML}/n_0^L$ gives account of the relative size of a precursor molecule and detached ligands, with $\alpha = 1$ meaning that both have the same size. The characteristic rates are defined as $v_{gas}^{ML} = J/n_0^{ML}$ (gas supply rate of precursor molecules), $v_{des}^{ML} = 1/\tau^{ML}$ and $v_{des}^L = 1/\tau^L$ (desorption rates of precursor molecules and detached ligands being inversely proportional to their respective residence times), and as $v_{dis}^{ML} = f\sigma^{ML}$ and $v_{dis}^L = f\sigma^L$ (dissociation rate of precursor molecules and detached ligands with their respective dissociation cross sections).

The initial coverage conditions at $t = 0$ before switching on the electron beam flux ($f = 0$) are for the detached ligand

$\theta^L(t = 0) = \theta_0^L = 0$ (eqn 6c)

as no detached adsorbed ligands exist initially. For the adsorbed precursor molecule, the initial value follows from the previous text describing the standard FEBID model:

$\theta^{ML}(t = 0) = \theta_0^{ML} = v_{gas}^{ML}/(v_{gas}^{ML} + v_{des}^{ML})$ (eqn 6d).

which follows from $\partial\theta^{ML}/\partial t\ (t = 0) = 0$ and is describing the steady state condition of precursor molecule adsorption without electron irradiation[26, 29, 31]. Eqn. 6d describes the highest coverage obtainable for ML upon adsorption. For non-steady state exposure strategies, it should be replaced by the actual (lower) coverage values [12]. For simplicity we stick to eqn. 6d without loss of generality.

**3.1 Coverages as function of dwell time**

In the case of pulsed exposure, the coverage is a function of dwell time $t_d$ during which the electron beam is on. The set of differential equations (6a&b) can be solved analytically and coverages $\theta^{ML}$ and $\theta^L$ for the precursor and the detached ligands, respectively, can be presented as sum of exponential functions:

$\theta^{ML}(t) = \theta_1 e^{\nu_+ t} + \theta_2 e^{\nu_- t} + \theta_\infty^{ML}$ (eqn. 7a)

$\theta^L(t) = -\theta_1 \frac{(\nu_+ + v_\Sigma^{ML})}{v_{gas}^{ML}} e^{\nu_+ t} - \theta_2 \frac{(\nu_- + v_\Sigma^{ML})}{v_{gas}^{ML}} e^{\nu_- t} + \theta_\infty^L$ (eqn. 7b)

The parameters in both equations are explained hereafter. The parameters $\theta_1, \theta_2$ are integration constants. The steady state coverage values under irradiation $\theta_\infty^{ML}$ and $\theta_\infty^L$ are derived from $\partial\theta^{ML}/\partial t = 0$ and $\partial\theta^L/\partial t = 0$ as:

$\theta_\infty^{ML} = \theta^{ML}(t \to \infty) = v_\Sigma^L v_{gas}^{ML}/(v_\Sigma^L v_\Sigma^{ML} + \alpha v_{gas}^{ML} v_{dis}^{ML}) = (v_{des}^L + v_{dis}^L)t_c$ (eqn. 7c)

$\theta_\infty^L = \theta^L(t \to \infty) = \alpha v_{dis}^{ML} v_{gas}^{ML}/(v_\Sigma^L v_\Sigma^{ML} + \alpha v_{gas}^{ML} v_{dis}^{ML}) = \alpha v_{dis}^{ML} t_c$ (eqn. 7d)

The sum rates $v_\Sigma$ for precursor molecules and detached ligands are defined as $v_\Sigma^{ML} = v_{gas}^{ML} + v_{des}^{ML} + v_{diss}^{ML}$ and $v_\Sigma^L = v_{des}^L + v_{diss}^L$, respectively. The expression $t_c = v_{gas}^{ML}/(v_\Sigma^L v_\Sigma^{ML} +$ $\alpha v_{gas}^{ML} v_{dis}^{ML}$) can be regarded as a characteristic time of the process involving both intact precursor and detached ligands. The characteristic reaction constants $v_+$ and $v_-$ of the ligand co-deposition model are derived as:

$$v_\pm = -\left(\frac{v_\Sigma^{ML}+v_\Sigma^L}{2}\right) \pm \frac{\sqrt{(v_\Sigma^{ML}-v_\Sigma^L)^2 - 4v_{gas}^{ML}\alpha v_{dis}^{ML}}}{2} \quad \text{(eqn. 7e)}$$

Note that values for the reaction constants $v_\pm < 0$ and result in $e^{-v_\pm \cdot t} \to 0$ for $t \to \infty$, i.e. steady state conditions.

An interesting observation is that the stationary coverage ratio of adsorbed precursor molecules to detached ligands derives from equations 7a&b as:

$$\frac{\theta_\infty^{ML}}{\theta_\infty^L} = (v_{des}^L + v_{dis}^L)/(\alpha v_{dis}^{ML}) \quad \text{(eqn. 8)}$$

and does not depend on the gas supply and desorption rate of the initial ML molecule.

### 3.2 Deposition rates and yields as function of dwell time

Two reaction yields need to be evaluated. Introducing the index $M$ for the deposited fragment containing the metal atom and the index $\Lambda$ for the co-deposited ligand residue the yield can be expressed in analogy to eqn. 3 as:

$$Y^M = \frac{Y_{max}^M}{t_d} \int_0^{t_d} \theta^{ML}(t)dt \quad (9a)$$

$$Y^\Lambda = \frac{Y_{max}^L}{t_d} \int_0^{t_d} \theta^L(t)dt \quad (9b)$$

with $Y_{max}^M = \sigma^{ML} n_0^{ML}$ and $Y_{max}^\Lambda = \sigma^L n_0^L$.

Integrating equations 9a&b gives for the time-averaged yields:

$$Y^M(t) = \frac{Y_{max}^M}{t_d}\left[\frac{\theta_1}{v_+}(e^{v_+ t}-1) + \frac{\theta_2}{v_-}(e^{v_- t}-1)\right] + Y_{max}^M \theta_\infty^{ML} \quad \text{(eqn. 10a)}$$

$$Y^\Lambda(t) = \frac{Y_{max}^\Lambda}{t_d}\left[\frac{\theta_1}{v_{gas}^{ML}}\left(1+\frac{v_\Sigma^{ML}}{v_+}\right)(e^{v_+ t}-1) + \frac{\theta_2}{v_{gas}^{ML}}\left(1+\frac{v_\Sigma^{ML}}{v_-}\right)(e^{v_- t}-1)\right] + Y_{max}^\Lambda \alpha \theta_\infty^L \quad \text{(eqn. 10b)}$$

Note that the maximum yields are related by $Y^\Lambda_{max} = Y^M_{max} v^L_{dis}/(\alpha v^{ML}_{dis})$. The terms $(e^{v_\pm t} - 1)/t$ approach zero for steady state conditions ($t \to \infty$) since $v_\pm < 0$, see eqn. 7e, and the steady state yields become:

$$Y^M_\infty = Y^M(t \to \infty) = Y^M_{max} \theta^{ML}_\infty = Y^M_{max}(v^L_{des} + v^L_{dis})t_c \quad \text{(eqn. 11a)}$$

$$Y^\Lambda_\infty = Y^\Lambda(t \to \infty) = Y^\Lambda_{max} \theta^L_\infty = Y^\Lambda_{max} \alpha v^{ML}_{dis} t_c = Y^M_{max} v^L_{dis} t_c \quad \text{(eqn. 11b)}$$

with the characteristic time $t_c$ defined for equations 7c&d.

The initial yields ($t \to 0$) derive directly from the initial coverage expressions in equations 6c&d:

$$Y^M_0 = Y^M(t \to 0) = Y^M_{max} \theta^{ML}_0 = Y^M_{max} \frac{v^{ML}_{gas}}{v^{ML}_{gas} + v^{ML}_{des}} \quad \text{(eqn. 12a)}$$

$$Y^\Lambda_0 = Y^\Lambda(t \to 0) = Y^\Lambda_{max} \theta^L_0 = 0 \quad \text{(eqn. 12b)}$$

Finally, the total deposition yield is given by $Y = Y^M + Y^\Lambda$. (eqn. 13)

### 3.3 Composition as function of dwell time & steady state

The key feature of the ligand co-deposition model presented in this work is that it can predict composition trends of the FEBID material with external parameters, when working with a single precursor. The predicted composition is $M_{(x^M)}\Lambda_{(x^\Lambda)}$, with $x^M$ and $x^\Lambda$ being the atomistic (molar) percentages of the deposited metal $M$ and ligand fragment $\Lambda$, respectively, and $x^M + x^\Lambda = 1$. The composition defined here follows directly the deposition yields discussed in section 3.2 since the atomistic percentages $x_M$ and $x_\Lambda$ describe the share of each deposition yield with respect to the total yield $Y^M + Y^\Lambda$

$$x^M(t) = \frac{Y^M(t)}{Y^M(t) + Y^\Lambda(t)} = \frac{1}{1 + Y^\Lambda(t)/Y^M(t)} \quad \text{(eqn. 14a)}$$

$$x^\Lambda(t) = \frac{Y^\Lambda(t)}{Y^M(t) + Y^\Lambda(t)} \quad \text{(eqn. 14b)}$$

The initial and steady state values for $x^M$ are as follows.

For electron exposure times $t \to 0$ equations 14a&b give

$$x^M_0 = x^M(t \to 0) = 1 \quad \text{(eqn. 14c)}$$

since $Y_0^\Lambda = 0$ holds for the initial condition. Hence, the highest metal content is achieved for shortest electron beam exposure times. This behavior is a direct consequence of the two dissociation reactions (reaction 2a & b) entering the model, see also Figure 1. Extension to other potential reactions influencing composition with electron dwell time will be discussed in section 5.

Inserting equations 11a&b into equation 14a yields for the metal content at steady state electron exposure conditions $t \to \infty$

$$x_\infty^M = x^M(t \to \infty) = 1/(1 + v_{dis}^L/(v_{des}^L + v_{dis}^L)) = \frac{1}{1+\frac{1}{1+\frac{v_{des}^L}{v_{dis}^L}}}. \quad \text{(eqn. 14d)}$$

When this model applies, the steady state value corresponds to the lowest metal content obtained in FEBID nanoprinting. Interestingly, the steady state composition of the FEBID nanoprint material is fully independent of the pristine precursor kinetics, i.e. the delivery, desorption, and e-dissociation behaviour of the metalorganic precursor molecule ML. The metal content is only dependent on the detached ligand kinetics, namely on the $v_{dis}^L/v_{des}^L$ ratio. Figure 5 graphically represents this functional dependence.

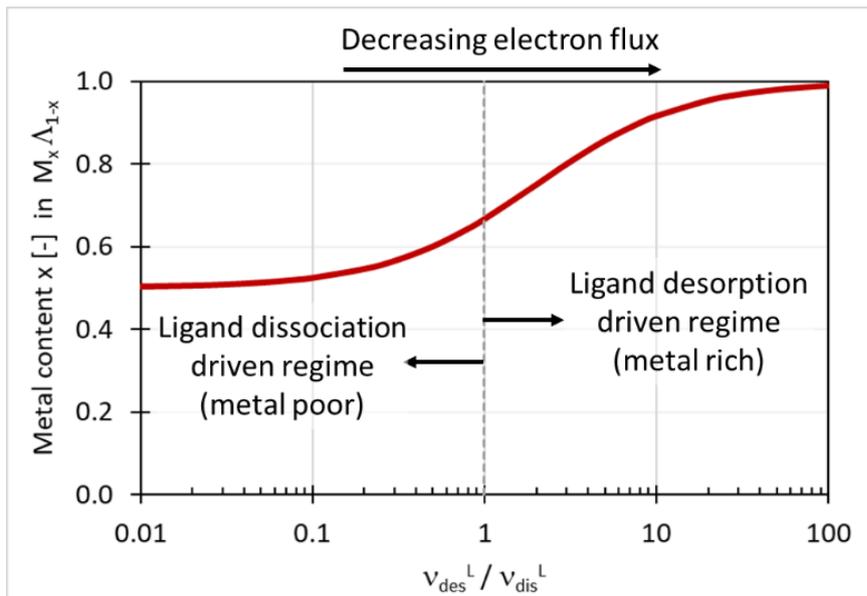

**Figure 2.** Steady state metal content in FEBID nanoprint material versus the detached ligand's rate ratio of desorption to dissociation according to equation 14d. Note that for real case

scenarios the asymptotic values for the metal content $1/2 \leq x \leq 1$ will scale as $\mu/(1 + \lambda) \leq x \leq \mu$ according to equation 15c.

For the ligand desorption-driven regime, $v_{des}^L \gg v_{dis}^L$, the metal content converges to $x_\infty^M = 1$ (100 at.%). In this case $x^M(t) = 1$ for all electron dwell times $t$. For the ligand dissociation driven regime, $v_{des}^L \ll v_{dis}^L$, the metal content converges to $x_\infty^M = 1/2$ (50 at.%). This is shown in figure 2. For the case $v_{dis}^L = v_{des}^L$ it follows that $x_\infty^M = 1/3$ ($\approx 67$ at.%). For now, equations 14c&d define the metal at.% range within $0.5 \leq x^M \leq 1$ as we considered the deposited reaction products of reactions 2 a&b consisting of one fragment M (pure metal atom) and one fragment $\Lambda$ (deposited residue of ligand *L*) and thus can be considered only as upper limit metal content predictions. In a real case scenario, the deposited M fragment may contain still some remaining ligand parts, hence a variable metal share $\mu$ ($0 < \mu \leq 1$) due to incomplete dissociation and the deposited ligand fragment $\Lambda$ often contains more than one atom $\lambda \geq 1$ ($\lambda \in \mathbb{N}$). To account for these scenarios the metal atomic percentage can be formulated as

$$x^M = \frac{\mu Y^M}{Y^M + \lambda Y^\Lambda} \quad \text{(eqn. 15a)}$$

with

$x_0^M = \mu$ (eqn. 15b)

and

$x_\infty^M = \mu/(1 + \lambda v_{dis}^L/(v_{des}^L + v_{dis}^L))$. (eqn. 15c)

For the ligand desorption driven regime, $v_{des}^L \gg v_{dis}^L$, $x_\infty^M \approx \mu$ and for the dissociation-driven ligand regime, $v_{dis}^L \gg v_{des}^L$, $x_\infty^M = \mu/(1 + \lambda)$.

For completeness, we should mention that both $\mu$ and $\lambda$ may potentially change with the electron flux. This discussion is beyond the scope of the present article.

## 4. Model assessment using literature data

In the following, we will discuss the electron pixel dwell time dependence of surface coverage, yields, and composition, predicted by our model, using the parameter set in table 1. The parameters such as cross section for dissociation and average desorption time have been chosen to be of the same order of magnitude as experimental values published in the literature [29, 33, 35, 36]. The parameters such as molecule flux and electron flux, which are controlled during experiments, were chosen based on typical values obtained in the experiments from our previous works, as well as available in literature [same sources as above, but this time also with our silver nanomaterials]. It turns out that the derived maximum deposition yield for the metal-containing fragment was close to one percent, so we set $Y_{max}^M = 1\%$. Note that both maximum yields are related via the choice of the dissociation rates as $Y_{max}^\Lambda = Y_{max}^M \cdot v_{dis}^L / (\alpha v_{dis}^{ML})$. The bottom part of Table 1 contains the initial and steady state values of surface coverage, yields, and composition, as derived from the model introduced in the previous sections.

**Table 1.** Parameters of precursor molecules and ligands used for calculations, together with their respective initial and steady state values. Note that the second set for the ligand only differs by a 100 times lower dissociation rate (dissociation cross section) from ligand set 1. The dwell time $t_i$ is an approximate indicator for the inflection point of the metal content curve.

| Parameter | Precursor (ML) | Ligand (L) set 1 | Ligand (L) set 2 |
|---|---|---|---|
| Precursor flux $J$ [1/(m²s)] | $2.0 \cdot 10^{20}$ | --- | --- |
| Adsorption site density $N_0$ [m⁻²] | $2.0 \cdot 10^{18}$ | $2.0 \cdot 10^{18}$ | $2.0 \cdot 10^{18}$ |
| **Gas supply rate** $v_{gas} = J/N_0$ [s⁻¹] | $1.0 \cdot 10^2$ | --- | --- |
| Electron flux $f$ [/(m²s)] | | $2.0 \cdot 10^{24}$ | |
| Cross section for dissociation $\sigma$ [m²] | $5.0 \cdot 10^{-21}$ | $5.0 \cdot 10^{-19}$ | $5.0 \cdot 10^{-21}$ |
| **Dissociation rate** $v_{dis} = \sigma f$ [s⁻¹] | $1.0 \cdot 10^4$ | $1.0 \cdot 10^6$ | $1.0 \cdot 10^4$ |
| Average desorption time $\tau$ [s] | $1.0 \cdot 10^{-3}$ | $1.0 \cdot 10^{-5}$ | $1.0 \cdot 10^{-5}$ |
| **Desorption rate** $v_{des} = 1/\tau$ [s⁻¹] | $1.0 \cdot 10^3$ | $1.0 \cdot 10^5$ | $1.0 \cdot 10^5$ |
| Parameters derived from analytical formulas | | | |
| $\alpha$ | | 1.0 | |
| $\theta_0$ | $9.1 \cdot 10^{-2}$ | 0.0 | 0.0 |
| $\theta_\infty$ | $9.0 \cdot 10^{-3}$ | $8.2 \cdot 10^{-5}$ | $8.2 \cdot 10^{-4}$ |
| $\theta_\infty^{ML}/\theta_\infty^L$ | | $1.1 \cdot 10^2$ | $1.1 \cdot 10^1$ |
| $Y_{max} = \sigma \cdot n_0$ | $1.0 \cdot 10^{-2}$ | 1.0 | $1.0 \cdot 10^{-2}$ |
| $Y_{max}^{\Lambda}/Y_{max}^{M}$ | | $1.0 \cdot 10^{2}$ | $1.0$ |
| $Y_0 = Y_{max} \cdot \theta_0$ | $9.1 \cdot 10^{-4}$ | $0.0$ | $0.0$ |
| $Y_{\infty} = Y_{max} \cdot \theta_{\infty}$ | $9.0 \cdot 10^{-5}$ | $8.2 \cdot 10^{-5}$ | $8.2 \cdot 10^{-6}$ |
| $x_0^M$ | | $1.0$ | $1.0$ |
| $x_{\infty}^M$ | | $5.2 \cdot 10^{-1}$ | $9.2 \cdot 10^{-1}$ |
| Dwell time $t_i$ [s] | | $9.1 \cdot 10^{-7}$ | $9.1 \cdot 10^{-6}$ |

**4.1 Dependence of surface concentrations of precursor molecules and ligands on exposure time**

Figure 3 presents the coverage of precursor and detached ligands as function of electron beam exposure time. The parameters used in these calculations are presented in Table 1. The data point legends in Figure 3 designate the characteristic time scales of gas supply, dissociation, and desorption for the precursor (purple) and the detached ligands (green). These are derived as inverse values of the corresponding rates of table 1, namely $1/\nu_{gas}^{ML}$, $1/\nu_{des}^{ML,L}$, and $1/\nu_{dis}^{ML,L}$. The datapoints designated as "sum" correspond to the characteristic times defined by $1/\nu_{\Sigma}^{ML}$ (precursor, purple) and $1/\nu_{\Sigma}^{L}$ (ligand, green), see eqns. 7 c&d.

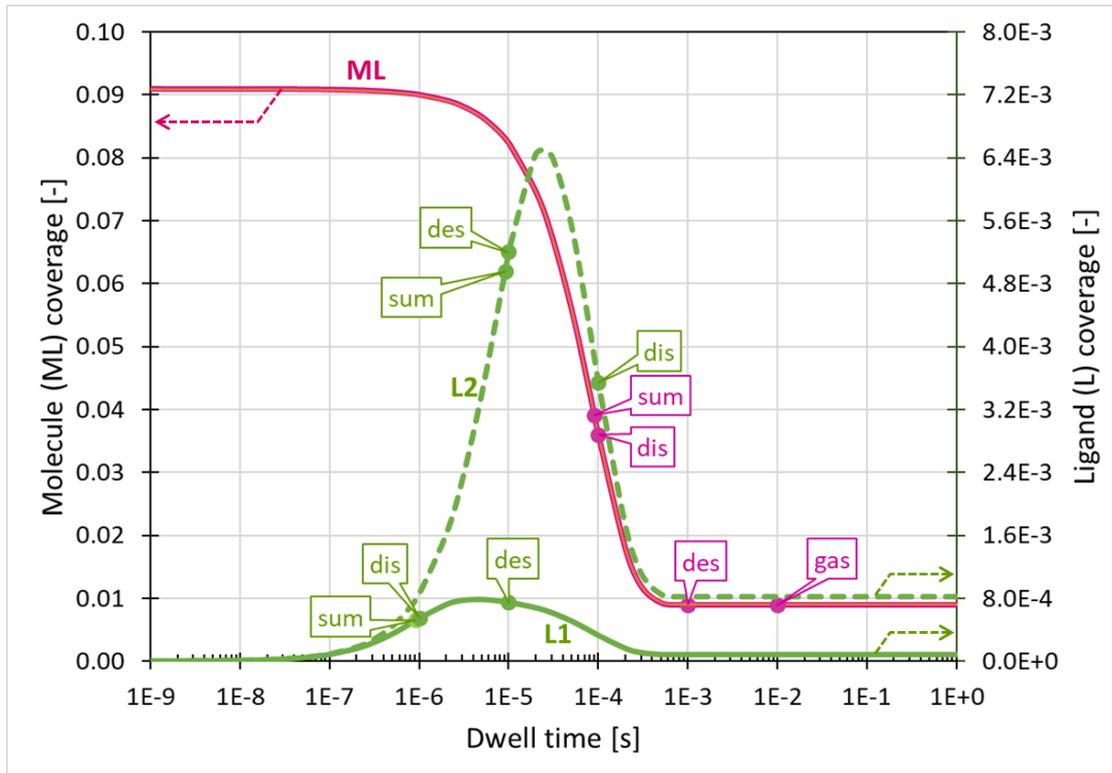

**Figure 3.** Evolution of surface coverage with electron exposure time of precursor molecules (purple curve) and detached ligands (green curves: full line set 1, dashed line set 2). For the explanation of data point legends, see text. Note that the initial and steady state values correspond to the values given in table 1. The coverages for the molecule ML do not change visibly for the two ligand sets. The data presented in this and following graphs were calculated for a continuous precursor supply.

The surface coverage of precursor molecules (purple) gradually decreases from an initial value of about 9% as exposure time increases, as molecules are increasingly dissociated by the dwelling electron beam. The point of inflection is around $t \approx 1/\nu_\Sigma^{ML}$ and evolves into the steady state value. The steady state value is reached at $t \approx 1/\nu_{des}^{ML}$. The surface coverage of the precursor molecules is not visibly changed by the two specific parameter sets chosen in table 1 for the ligands. The surface coverage of detached ligands (green) rises with increasing exposure time, as these are formed from dissociated precursor molecules. It reaches a maximum value, and then decreases due to molecule depletion, to the corresponding steady state value. The first inflection point during the rise of the ligand coverage falls approximately into the time range $t \approx 1/\nu_\Sigma^{L}$. The time position of the maximum ligand coverage does not follow a simple combination of the characteristic time scales, while the steady state onset is naturally close to the one for the precursor molecule. The higher peak for the detached ligand set 2 is simply due to the lower dissociation (depletion) rate with respect to ligand set 1. Please note that the shape of the graphs, as well as exact coverage values for both precursor and detached ligands are specific to the parameters in table 1, hence, to the specific precursor and the experimental conditions (electron flux and temperature determining the desorption rate) in FEBID nanoprinting. Supporting Information section S1 shows additional information on this behavior.

**4.2 Dwell time dependence of deposition yields of metal and ligands**

Figure 4 presents the deposition yields for the metal, the ligand residue, and the sum of both, using again the data of table 1. As for the previous figure 3, the datapoints designated as "sum" correspond to the characteristic times defined by $1/\nu_\Sigma^{ML}$ (precursor, purple) and $1/\nu_\Sigma^{L}$ (ligand, green).

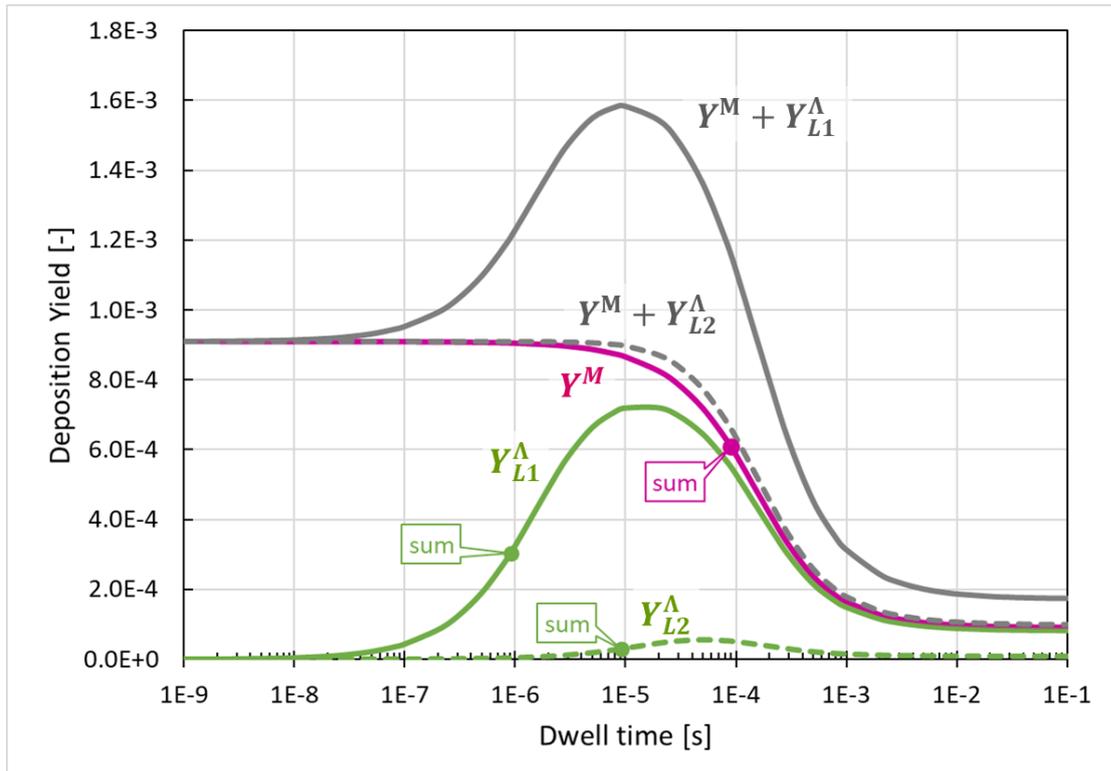

**Figure 4.** Evolution of deposition yields with dwell time of metal containing part (purple curve), ligand residue (green curves: full line set 1, dashed line set 2), and total yield (grey curves). For the explanation of data point legends, see text. Note that the initial and steady state values correspond to the values given in table 1. The yields for the metal do not change visibly for the two ligand sets.

The metal yield (purple) gradually decreases from about 0.9% of its maximum value with increasing exposure time, as precursor molecules are increasingly dissociated by the dwelling electron beam. The point of inflection corresponds to $t \approx 1/v_{\Sigma}^{ML}$, followed by the evolution into the steady state value. As previously noted, the metal yield is not visibly changed by the two specific ligand parameter sets chosen in table 1. The ligand residue yield (green) shows first a rise with increasing exposure time, as more detached ligands become available for dissociation, see fig. 3. As the electron exposure time increases, the yield reaches a maximum value, and then decreases due to detached ligand depletion, eventually approaching a steady state value. The depletion of the ligands is caused by the depletion of pristine precursor molecules for longer dwell times. The first inflection point during the rise of the ligand residue yield is approximately described by $t \approx 1/v_{\Sigma}^{L}$, and the steady state onset is naturally close to the one for the precursor. The higher yield peak for the ligand set 1 is due to the higher dissociation rate (leading to more ligand residue deposition) with respect to ligand set 2.

In general, all steady state yield values are reached in a later exposure time than for the corresponding coverages discussed previously. This is due to the integration over the exposure time, see eqns. 10 and 12. The total yield curves for the two ligand sets have different shapes. Keeping in mind that the deposition rate (and hence the thickness of the deposit) is proportional to the deposition yield, see eqn. 5, in an experiment one would measure a peak value for ligand set 1 with increasing dwell time due to the considerable contribution of ligand fragment co-deposition. Such a peak in the growth rate (yield) curve is a specific feature of ligand co-deposition and not predictable with single species FEBID models [26]. For ligand set 2, the overall ligand residue co-deposition is less pronounced, such that no peak occurs in the total yield, and one would measure a steady decay of growth rate with increasing dwell time until the steady state value. As for the previous section, we note that the shape of the graphs as well as exact values for both metal and ligand fragment yields are specific to the parameters in table 1. Figure S2 in the Supporting Information shows additional information on this behavior and the following general trends are observed:

(i) The metal deposition yield continuously decreases with increasing exposure time as the pristine precursor molecule is naturally depleted (adsorbate limited FEBID regime due to limited continuous molecule transport). The lowest and highest values correspond to $Y_\infty^M$ and $Y_0^M$ given in eqns. 11a and 12a.

(ii) The inflection point for the metal yield curve occurs at the exposure time $t_i \approx 1/\nu_\Sigma^{ML}$.

(iii) The ligand residue yield increases with increasing exposure time to a value given by $Y_\infty^\Lambda$.

(iv) Under conditions where the total deposition yield shows a peak, the inflection points are located at dwell times $t_i \approx 1/v_\Sigma^L$ (ascending) and $t_i \approx 1/v_\Sigma^{ML}$ (descending).

(v) If on the contrary, the total deposition yield is monotonous with exposure time, the only descending inflection point is located at dwell time $t_i \approx 1/v_\Sigma^L$ or $t_i \approx 1/v_\Sigma^{ML}$, see Supporting Information, Figure S2.

(vi) The total deposition yield can either decrease, increase monotonously, or exhibit a peak, as a function of the exposure time. Notably, a monotonically increasing trend or a peak in the total deposition yield uniquely indicates the presence of ligand co-deposition. In contrast, models assuming instantaneous ligand desorption consistently predict a decrease in total deposition yield with increased exposure time. For the case, where the ligand co-deposition is negligible, the total deposition yield collapses to the FEBID standard deposition model summarized in section 1, see Figure S3 in Supporting Information.

**4.3 Exposure time dependence of deposit composition**

Despite the previously discussed variety of the shapes of deposition yield curves, Figure S2 in the Supporting Information also shows that the metal content $x^M$ of the FEB nanoprinted material follows one single general shape. Due to the assumptions of the model, the composition always starts with its maximum value $x_0^M = 1$ as initial value (or $x_0^M = \mu$ in case of incomplete dissociation) and decreases to the respective steady state values derived in section 3.3. For the ease of the following discussion, we chose $\mu = \lambda = 1$, see equations 15. From the parameter set in table 1 we obtain for the two ligand sets 1 and 2: $x_\infty^M = 0.52$. and $x_\infty^M = 0.92$, respectively. Figure 5 and Figure S2 in the Supporting Information show that the metal content decreases monotonically with increasing electron exposure time, from $x_0^M$ to $x_\infty^M$ ($x_0^M > x_\infty^M$), passing through an inflection point where the change is most rapid.

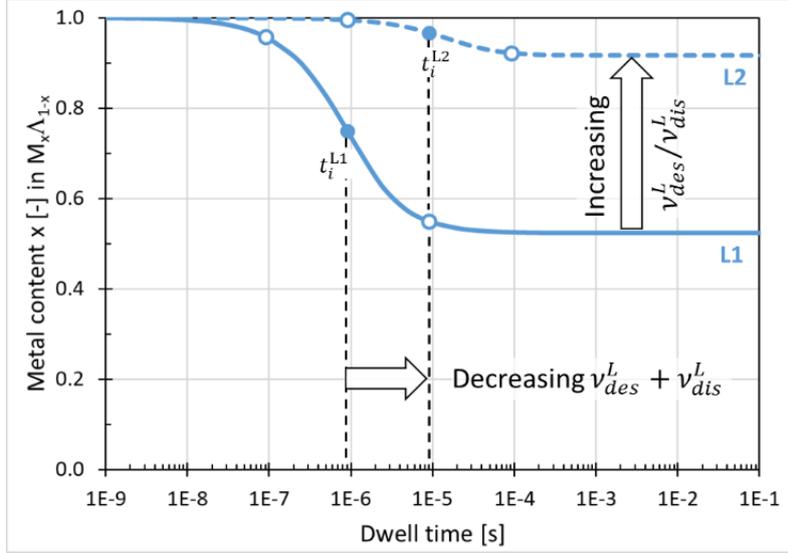

**Figure 5.** Evolution of deposit composition with electron exposure time for the two detached ligand sets (full line set 1, dashed line set 2) in table 1. The solid circles denote the positions of $t_i$ which approximately define the dwell time at inflection of $x^M$, see eqn. 16. The empty circles mark dwell times of $0.1t_i$ and $10t_i$. Note that the initial and steady state values correspond to values given in table 1 and that $\mu$ and $\lambda$ were set to 1 (see section 3.3) to show the unobscured functional behavior of the metal content.

From the graphical inspection of several precursor molecule & ligand data sets (see Supporting Information section S2), we identified that the electron exposure time at the point of inflection $x_i^M$ is approximately:

$t_i \approx 1/v_\Sigma^L = 1/(v_{des}^L + v_{dis}^L)$ . (eqn 16a)

Furthermore, for small electron exposure times

$t \leq 0.1 t_i = 0.1/v_\Sigma^L : x_0^M \leq x_0^M \leq 0.9 x_0^M$ (eqn 16b)

the metal content stays close to its initial value $x_0^M = 1$, or more generalized to $x_0^M = \mu$ (see section 3.3.), within approximately 10%. For electron exposure times

$t \geq 10 t_i = 10/v_\Sigma^L : x_\infty^M \leq x^M < 1.1 x_\infty^M$. (eqn 16c)

the metal content stays close to the steady state value $x_\infty^M = 0.5$, or more generalized $x_\infty^M = \mu/(1 + \lambda v_{dis}^L/(v_{des}^L + v_{dis}^L))$ (see section 3.3) within approximately 10%.

We refrain from a rigid mathematical derivation of approximations 16a-c involving lengthy expressions that can be solved only numerically. Exact values can be evaluated by the readers inserting rates of their specific interest into equations 15a&b.

Figure 5 highlights the key prediction the ligand co-deposition model is able to deliver concerning the metal composition of a FEBID deposit: both the dwell time onsets of the metal content change and the metal content for continuous exposure (steady state, dwell time > 100 $t_i$), depend solely on the ligand's dissociation $v_{dis}^L$ and desorption $v_{des}^L$ values. This suggests that under the conditions where this model applies, the final composition of a FEBID material is not governed by the surface kinetics or deposition regime of the pristine precursor molecule, but rather by the kinetics of the detached ligand.

## 5 Model assessment using experimental data
### 5.1 Growth rate comparison

In order to investigate the applicability of the model to real life FEBID processes, the following experiment has been performed: a series of pulsed-exposure spot deposits were prepared, each with different dwell times, ranging from 50 ns up to 0.1 s. All deposits were designed to have the same total exposure time of 0.1 s and the same current of 21 pA, giving as a result the same dose. The number of loops varied to maintain the constant total exposure time. The acceleration voltage was 3 kV. The refresh time was set to 1 ms, to allow a good replenishment of the precursor molecules in-between pulses. A Zeiss prototype mask repair tool was used to prepare the deposits, using $Cr(CO)_6$ as precursor, which was delivered through a built-in gas injection system [37]. After deposition, the heights of the spots were measured using a NT-MDT AFM – Raman system, with Bruker RTESPA 300 tips having a nominal tip diameter of 8 nm. By experimentally measuring the height of the deposits and taking into account the total exposure time, the growth rate in nm/s of each spot was calculated and plotted as a function of applied dwell time (see Figure 6 - black diamonds). Growth rate calculations using the ligand co-deposition model were performed with the parameters shown in Table 2. The parameter set was selected to achieve visual agreement with the experimental data.

**Table 2:** Parameters used for the calculated growth rates in figure 6.

| Parameter | Precursor (ML) | Ligand (L) |
|---|---|---|
| Precursor molecule flux J [#/(m²s)] | $4.0 \cdot 10^{20}$ | --- |
| Adsorption site density $N_0$ [m⁻²] | $2.0 \cdot 10^{18}$ | $2.0 \cdot 10^{18}$ |
| **Gas supply rate [s⁻¹]** | $2.4 \cdot 10^{2}$ | -- |
| Electron flux [#e/(m²s)] | $2.0 \cdot 10^{24}$ | |
| Cross section for dissociation [m²] | $3.8 \cdot 10^{-20}$ | $6.0 \cdot 10^{-18}$ |
| **Dissociation rate [s⁻¹]** | $7.5 \cdot 10^{4}$ | $1.2 \cdot 10^{7}$ |
| Average desorption time [s] | $1.0 \cdot 10^{-4}$ | $1.0 \cdot 10^{-6}$ |
| **Desorption rate [s⁻¹]** | $1.0 \cdot 10^{4}$ | $1.0 \cdot 10^{6}$ |
| M volume [m³] | $1.8 \cdot 10^{-29}$ | |
| Λ volume [m³] | | $6.8 \cdot 10^{-29}$ |
| $V^M * f$ [nm/s] | $3.6 \cdot 10^{4}$ | |
| $V^\Lambda * f$ [nm/s] | | $1.4 \cdot 10^{5}$ |
| **Derived parameters from analytical formulas** | | |
| $\alpha$ | 1.0 | |
| $\theta_0$ | $2.3 \cdot 10^{-2}$ | 0.0 |
| $\theta_\infty$ | $2.8 \cdot 10^{-3}$ | $1.6 \cdot 10^{-5}$ |
| $\theta_\infty^{ML}/\theta_\infty^{L}$ | | $1.7 \cdot 10^{2}$ |
| $Y_{max} = \sigma \cdot n_0$ | $7.5 \cdot 10^{-2}$ | $1.2 \cdot 10^{1}$ |
| $Y_{max}^\Lambda/Y_{max}^M$ | | $1.6 \cdot 10^{2}$ |
| $Y_0 = Y_{max} \cdot \theta_0$ | $1.8 \cdot 10^{-3}$ | 0.0 |
| $Y_\infty = Y_{max} \cdot \theta_\infty$ | $2.1 \cdot 10^{-4}$ | $1.9 \cdot 10^{-4}$ |
| $R_{max} = V \cdot f \cdot Y_{max} \left[\frac{nm}{s}\right]$ | $2.7 \cdot 10^{3}$ | $1.6 \cdot 10^{6}$ |
| $R_{max}^\Lambda/R_{max}^M$ | | $6.0 \cdot 10^{2}$ |
| $R_0 = R_{max} \cdot \theta_0 \left[\frac{nm}{s}\right]$ | $6.3 \cdot 10^{1}$ | 0.0 |
| $R_\infty = R_{max} \cdot \theta_\infty \left[\frac{nm}{s}\right]$ | 7.6 | $2.6 \cdot 10^{1}$ |

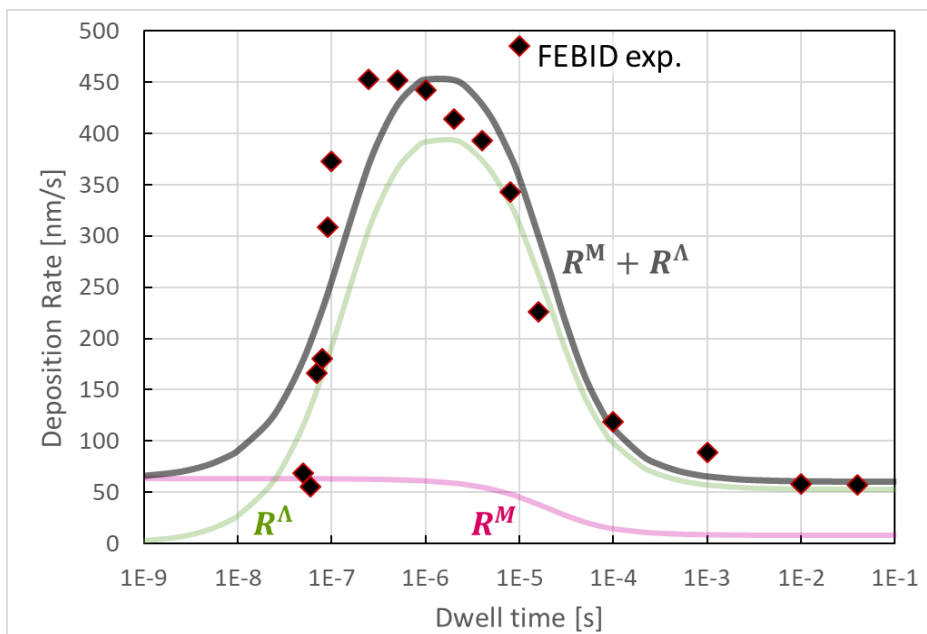

**Figure 6**. Comparison of experimentally observed growth rates (black diamonds) with the ligand co-deposition model (grey – total growth rate, purple – metal growth rate, green – ligand residue growth rate). Note that the model can capture the peak of the observed experimental growth rate.

The experimentally observed growth rate has a pronounced peak shape as a function of electron beam exposure time, which can be reproduced with the ligand co-deposition model. As previously remarked, this is a feature that previous models with instantaneous ligand desorption cannot predict. Instead, these models lead to a monotonously decaying growth rate with increasing electron exposure time due to the depletion (dissociation) of the adsorbed molecules. Incorporating the surface kinetics of the detached ligand, as done in the ligand co-deposition model, allows for the simulation of both metalorganic precursor deposition (purple curve in Figure 6, showing depletion with increasing exposure time) and ligand deposition (green curve in Figure 6, showing a peak-shaped behavior due to continuous supply from the dissociation of metalorganic molecules). Growth rate peak investigations at such dwell time range have not yet been reported in literature, likely due to technical limitations. Notably, the maximum of the peak appears around dwell times of 100–200 ns, a dwell time range which is difficult to access using standard scanning electron microscopes. Typical dwell time ranges used in literature for 3D growth vary from single microseconds to several milliseconds [12, 22, 25, 33].

## 5.2 Metal content predictions

The co-deposition model can be applied to predict the spatial trends of metal content of the deposited material at given deposition conditions. In our previously published FEBID results using silver carboxylates [32, 33, 38], steady state spot deposits obtained within the primary electron beam were surrounded by a halo deposit, see an example in Figure 7a. Halo deposits are due to interactions between adsorbed precursor molecules with backscattered and secondary electrons of second type (generated through interaction between backscattered electrons and the substrate) both exiting the surface at a distance with respect to the primary electron beam [32]. Energy dispersive X-ray spectrometry (EDS) measurements revealed that the silver content in the halo region was usually higher than in the deposit center. As an example, we mention FEBID experiments with the precursor Ag-(2,2- dimethylbutanoato-κO)- (AgO$_2$Me$_2$Bu) [32]; the centre versus halo EDS measurements resulted in silver metal content of $x^M$ = 42at.% vs 70 at.%, respectively, the remainder being mainly carbon and a minor oxygen content.

To access the spatial electron flux distribution we performed Monte Carlo simulations with the CASINO software [39] which generally follow the trend shown in figure 7b (blue dashed curve): the flux of the electrons in the halo region is significantly lower than in the center, where the primary beam impinges. The BSE distribution was simulated with the beam parameters of 20 kV acceleration voltage, 0.5-0.6 nA electron beam current, and a beam with a 220 nm FWHM. This simulated electron flux distribution $f(r)$ is then applied to the co-deposition model at steady state (our continuous spot exposures lasted >1s), to obtain changes of metal content as a function of the distance from the beam center. The steady state expression of the metal content from section 3.3 as function of distance $r$ from the electron beam center becomes in this case

$$x_\infty^M(r) = \frac{\mu}{1+\frac{\lambda}{1+\frac{1}{\tau^L \cdot \sigma^L f(r)}}}, \quad \text{eqn 17}$$

The maximum metal content which can be obtained in steady state conditions is when the condition $\sigma^L \tau^L f(r) \ll 1$ holds, namely where the electron flux is lower than $f(r) \ll (\sigma^L \tau^L)^{-1}$. This relation is satisfied at the peripherical halo region where $f(r) \to 0$ and the maximum metal content gets $x_\infty^M(\sigma^L \tau^L f(r) \ll 1) \cong \mu$. In view of the several orders of magnitude increase of the electron flux value $f(r)$ towards the centre region we expect the relation to change to $f(r) = (\sigma^L \tau^L)^{-1}$ at a certain radius $r$ where the metal content would have already decreased accordingly to eqn. 17. If the peak electron flux at the centre satisfies the condition $\sigma^L \tau^L f(r) \gg 1$, then the lowest metal content obtained here is $x_\infty^M(\sigma^L \tau^L f(r) \gg 1) \cong \frac{\mu}{1+\lambda}$. The boundary composition values are the same as derived in section 3.3. However, here we derived their spatial arrangement in steady state spot deposits.

An example of such a spot deposit is presented in Figure 7a and was obtained with silver pivalate $Ag_2[\mu - O_2C^tBu]_2$,). The centre, where the beam was position and the halo regions are marked with blue and red circles. The with dashed-line circles marks he spots where the composition was measured [32]. The spatial distributions of metal content in fig. 7b were calculated with the factors $\lambda = \mu = 1$ to show the unobscured trend on the $\tau^L \cdot \sigma^L$ parameter. Changing their value would allow to adjust to the experimentally determined metal content.

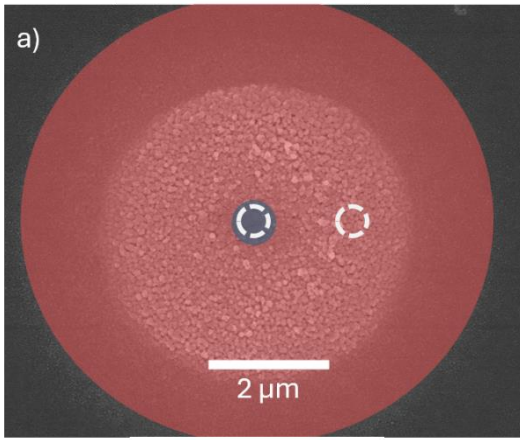
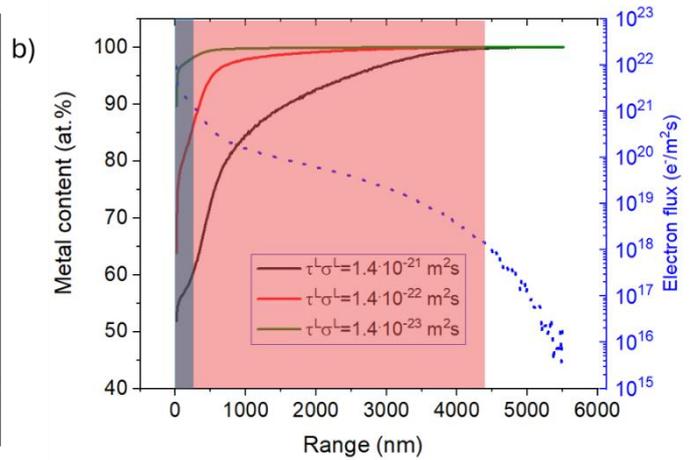

**Figure 7.** a) Continuous exposure spot deposit prepared with $Ag_2[\mu - O_2C^tBu]_2$. The blue and red regions symbolize the place where the beam and the halo were positioned, respectively. The corresponding metal contents were 40 at.% (centre) and 60 at.% (halo), measured by EDS (white, dashed line circles) [32]. b) Calculated metal content as a function of distance from the beam center (Range = 0) for varying $\tau_L \sigma_L$. The electron flux profile originating from the primary and backscattered electrons was calculated using the simulation software CASINO (blue dashed line). The shaded regions correspond to (a) to indicate halo and primary beam regions.

As can be noticed, the metal content increases with the distance from the deposit center for all calculated curves and thus catches the experimentally observed trend. The calculated metal content curves result in a varying shape depending on the $\tau_L \sigma_L$ term. The higher the dissociation cross section $\sigma^L$ and the higher the residence time $\tau^L$ of ligands the more of them will be co-deposited. This is decreasing the metal content in absolute value in the center part of the profile $x^M(r)$ and forcing the profile to reach its maximum metal content at closer vicinity to the beam center. From the halo metal content in figure 7b it follows that $\mu = 0.60$. An estimation of the number of ligand atoms $\lambda$ from the center 40 at.% metal content is not possible since the peak electron flux at the centre does not satisfy the condition $\sigma^L \tau^L f(r) \gg 1$.

### 5.3 Comparison with other results reported in literature

Comparison of the model to the values reported in literature is challenging, as there are only a limited number of systematic studies covering the influence of FEBID parameters on the purity of deposited material. However, there are notable examples which can be confronted with our findings. First, the aforementioned dependance of the purity of the material as a function of the distance from the beam center has been observed for a handful of silver carboxylates [33], and gold precursors [37]. Importantly, for both silver and gold complexes, there is one ligand per one metal atom, as in the assumptions of the presented model. Second, we can also check our model for the case of $W(CO)_6$, which is one of the most widely used precursors, due to its excellent properties for FEBID and the possibility to 3D print superconducting nanomaterials [40] . Higher metal content in the FEB deposited material was obtained for shorter dwell times in [41], in good agreement with our model. However, for completeness we also mention another study reporting the opposite trend, i.e. low resistivity deposits due to high metal content or better percolation obtained for long exposure times [42]. However, the percolation mechanism for electrical conductivity may be dependent on the exposure dose and might obscure the ligand co-deposition trends.

Third, an indirect support of our model results from the systematic study of six precursors: $W(CO)_6$, $MeCpPtMe_3$, $Co_2(CO)_8$, $Co(CO)_3NO$, TEOS and $Me_2Auacac$ [43]. In this study, the FEBID growth temperature was varied by heating the substrate stage. For four precursors, the increase of FEBID growth temperature resulted in higher metal content compared to room temperature deposits. This is in line with our model findings. Temperature increase promotes the desorption of both ligands and precursor molecules from the surface via the well-known Arrhenius relation. For long exposure times, a higher ligand desorption rate leads to a higher metal content according to eqns. 14 and 17. But even at lower dwell times, a similar effect of temperature on purity is expected. For most of the compounds in this study, the ligands are short, formed by molecules like CO or NO. It is thus reasonable to assume that their desorption rate is higher than that of a pristine precursor molecule, hence increasing the metal content of a deposited material as temperature increases. For completeness, we mention that two precursors, $MeCpPtMe_3$ and TEOS, showed a negligible change of composition with temperature.

Fourth, $Co_2(CO)_8$ and iron carbonyls represent counter examples to our model. Since they are widely used for 3D FEBID nanomagnetic studies [5, 44-47], we would like to point out the reasons. In the case of continuous exposure with long dwell times to deposit pillar geometries [6, 9, 14, 45, 48], growing the structures under mass transport limited regime for the precursor molecules has shown to result in the highest purities [45]. This contrasts with the results of our model, which predicts that for continuous exposure, metal content depends solely on the ratio of desorption to dissociation rate of the ligands and is highest for short dwell times. This deviation can be explained by additional mechanisms this molecule is known for, and which were not included in the present model: $Co_2(CO)_8$ may adsorb in multilayers [29]; it may also present an autocatalytic deposition contribution [48, 49], with the molecule spontaneously decomposing on clean cobalt material. We also expect iron carbonyls to deviate from our model as autocatalysis was reported for $Fe_2(CO)_9$ [45] and $Fe(CO)_5$ [17].

## 6. Conclusions and outlook

As outlined in the previous sections, the ligand co-deposition model demonstrated its capability to capture trends related to composition and growth rate observed for several FEBID processes. We turn now to a discussion of the consequences of model simplifications and outline potential directions for future model extensions.

a) In the present study, the model does not include surface diffusion of precursor molecules and ligands. Surface diffusion and its influence in FEBID have been studied in various works, e.g. [25, 26, 31, 50]. In these works, it was shown that surface diffusion influences the shape and size of the deposit. However, it can be neglected for very short (<1 μs) or very long (>100 μs) dwell times (these numeric values being specific for the gas and electron exposure parameters involved in the related experiments). Surface diffusion is thus not expected to change drastically the conclusion about initial and steady state values drawn previously on coverage, yield, and composition. However, in the remaining dwell time range concentration gradients will induce surface flux of precursor molecules towards the irradiation center (due to their depletion) and ligand flux towards the periphery (to dilute away from the irradiation center). This tendency would increase the metal content in the central region. Surface diffusion can be accounted for by adding a characteristic rate for diffusion, for both precursor molecules and ligands. However, this will likely always require a fully numerical approach, at the cost of analytical solutions which more transparently show the propagation of parameters to the outcome. Our model is based on the Langmuir adsorption framework, which assumes that only a monolayer of precursor molecules can adsorb on the surface. Historically, the Langmuir approach has been applied to the FEBID continuum model as a pragmatic solution to the lack of adsorption-type data for FEBID-relevant molecules, while still yielding reasonably good agreement between experiments and theoretical predictions of deposit morphology [26, 31]. More recently, multilayer adsorption was incorporated in the work of Sanz Hernández and Fernández-Pacheco for Co$_2$(CO)$_8$ [29]. However, this extension is mathematically more complex and not easily solvable analytically. If adsorption-type data for FEBID precursors and surfaces were to become available, the model could be refined in the future to account for such effects.

b) Our model considers two electron induced dissociation reactions resulting in a deposited metal atom and the ligand (reaction 2a) and a deposited ligand fragment (reaction 2b), see also figure 1. The parameters $\mu$ (metal share, to account for incomplete dissociation) and $\lambda$ (number of ligand atoms, to account for the experimentally measured compositions containing more than one ligand atom) were introduced in equations 15 retroactively, after the original derivation. While this model setup can predict the fate of the ligand (co-deposition versus desorption) and the related trends, it does not proactively capture the additional complexity introduced by potential further dissociation reactions (pathways). To illustrate the diversity of possible pathways, let us consider a generic metal carbonyl with stoichiometry $M(CO)_n$. This molecule can potentially dissociate according to various reaction series when reacting with one electron: (i) $M(CO)_n + e^- \rightarrow M(CO)_{n-m} + mCO \uparrow$, where $m \leq n$, $(n, m \, \epsilon \, \mathbb{N})$ leading to CO loss, (ii) $M(CO)_n + e^- \rightarrow MC(CO)_{n-1} + O \uparrow$ leading to metal carbide formation, (iii) $M(CO)_n + e^- \rightarrow MO(CO)_{n-1} + C \downarrow$ forming metal oxide in a carbonaceous matrix, and (iv) $M(CO)_n + e^- \rightarrow M(CO)_{n-1} + C \downarrow + O \uparrow$ forming metal in a carbon matrix [51, 52]. When using $\mu$ and $\lambda$ to adjust the model to match the experimental metal content in a FEBID process, we assume that a single dominant reaction pathway governs the process. Under these conditions, the simplifying assumptions of our model appear to be valid. However, when dissociation pathways are interdependent and multiple pathways influence the process outcome, more sophisticated models involving several sets of differential equations will be required. Such models have been developed for electron beam induced etching [26, 27, 53] and adapting their formalism to FEBID may catch trends of inverse relationship between metal content and electron exposure time discussed in section 5.3. However, these approaches require providing parameters for each additional reaction. In contrast, our ligand co-deposition model requires a minimal set of input parameters.

c) Although the model developed here was based on single-ligand compounds, certain conclusions can be extended to multi-ligand compounds. This claim is supported by the fact that the experimental growth rate versus dwell time curve presented in section 5.1, obtained for the multi-ligand compound $Cr(CO)_6$, was in qualitative agreement with our predictions. The more ligands detach from the central metal atom upon electron irradiation, the more adsorption sites they block on the surface for incoming precursor molecules. This follows directly from eqn. 6a, where high ligand coverage limits the coverage of precursor molecules. The overall probability of detached ligand deposition should generally increase with the number of ligands $n$ in the pristine molecule following $1 - (1-p)^n$, where $p$ is the probability of one ligand being deposited and assuming that each ligand dissociates independently of the others. This simple probabilistic argument implies that a higher number of detached ligands on the surface increases the likelihood of detached ligand deposition, consistent with FEBID experiments using carbonyl precursors. For example, Fe, Co and FeCo carbonyls (with 3-5 ligands per metal atom) generally yield high purity deposits (from 80at.% of metal up to almost pure metal), whereas for W and Mo carbonyls, which have 6 ligands per molecule, generally result in lower metal content [20]. In the future, our model can be extended to explicitly treat multi-ligand precursors by generalizing eqn. 6a and adding more equations to describe coverages of different ligands. However, this extension is beyond the scope of the present study.

d) The FEBID ligand co-deposition model allows drawing a few conclusions for the design of FEBID nanoprinting conditions. For building 3D nanostructures, for example by using the f3ast package being based on the standard FEBID continuum model [25], it would be desirable to have approximately constant growth rate and (high) metal content of the deposited material over a large range of dwell times. This is particularly important for the longest dwell time regimes normally used, between tens of microseconds up to tens of milliseconds. To achieve such behaviour the trends shown in Figure 5 and derived in eqns 15c and 16c should be implemented for a hypothetical precursor. The yield and composition vs electron dwell time for such a precursor are shown as example in Figure 8. The exact parameters used in the calculations are presented in Table S3 in SI.

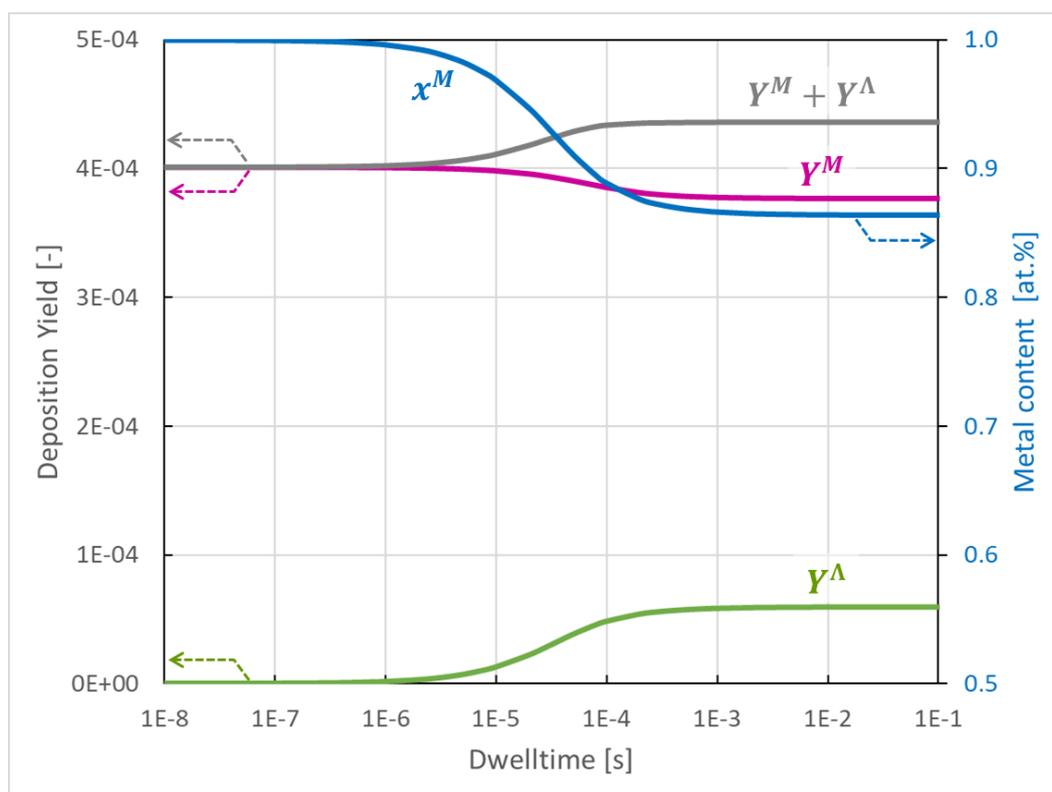

**Figure 8**. Example calculation of yields and metal content for a hypothetical precursor with a desirable performance for 3D nanoprinting: stable growth rate combined with high metal content across two large dwell time windows (10 ns $\leq t_d \leq$ 1 μs and $t_d \leq$ 1 ms). The involved parameters used for the calculation are included in the Supporting Information, see table S3.

This example highlights the important role of desorption of the detached ligands to achieve high purity deposits. Keeping a high desorption rate for detached ligands from the surface allows to have the high metal contents even at longer dwell times. This can be achieved, for example, by operating at high stage temperatures, though this would also naturally increase the desorption of precursor molecules. However, since the ligands in FEBID precursor molecules are typically chosen for their high volatility, increasing the stage temperature would have a greater effect on ligand desorption than on the precursor molecules themselves. On the precursor design side, the requirement of having a desirable large desorption rate suggests that the ligand should possess a high energy threshold to strongly adsorb on the surface once detached from the centre metal atom.

e) In recent years there has been a discussion in the FEBID community regarding the role of ligand length and size in achieving high purity deposits. Some people pointed out the cases of successful deposition of highly pure (> 90 at.%) Fe and Co FEBID deposits using short-ligand carbonyls [6, 14, 17, 45], and pure Au deposits achieved with $(PF_3)AuCl$ [15] and $Au(CO)Cl$ [16]. However, relatively high purities (on the level of 50- 75at%.) obtained with silver carboxylates with much longer carbon chains shed a new light in this realm. Studies on bis-ketoesterates of Cu and Pd [54, 55] show a high ligand removal rate from the pristine precursor of up to 90% (corresponding to 30 at.% metal content in the deposit) although the (*tert*-butylacetoacetate) ligand contains a large number of 8 C, 3 O, and 13 H atoms. From our model, it can be concluded that the quantities which should be considered are the ligand desorption and dissociation rates instead of the size of the ligands. Ligands with low dissociation and high desorption rate will be negligibly deposited under typical FEBID conditions, despite their size. This underlines the importance of further studies regarding the cross section for electron-induced dissociation and the average desorption time of ligands. It is worth noting that some ligands, even when their volatility upon cleavage from metal atom is not very high, can still efficiently desorb from the surface by forming more volatile species reacting on the surface. One such example for the silver carboxylate Ag (I) 2,2-dimethylbutanoate was described by Martinovic et al. [56], where an alkyl radical reacted to an alkene by hydrogen loss. For now, such a mechanism is not considered in our model, as the ligand is dissociated into volatile and non-volatile fragments upon irradiation and no further changes to the non-volatile fragment are considered. Such mechanisms can be integrated in the future by extending the set of surfaces reactions.

f) In section 5.2, we discussed the spatial distribution of metal content in spot deposits as a function of the spatial electron flux distribution. In many applications, such halo deposits are undesirable because they limit spatial resolution, can lead to unintended electrical pathways, and may largely affect a 3D nano-printing process. These halos can be often minimized by using small electron energies in FEBID. However, as demonstrated here, halo formation can be turned into a powerful analytical tool: it enables the investigation of ligand kinetics over a wide range of electron flux values. Within a single spot exposure, reactions 2a & b are effectively screened for deposited metal content and function of electrons available as surface reaction partners. Since the composition can be spatially mapped by spectroscopic methods such as energy or wavelength dispersive X-ray spectroscopy, this type of experimental data can support deterministic fitting to extract precursor molecule and ligand kinetic parameters. This has been exemplified with two spatially varying fluxes of chemical compounds at varying growth temperatures [57].

## 7. Summary

The FEBID co-deposition model presented in this work is a new theoretical tool for describing and analyzing surface processes during focused electron beam induced deposition. Unlike the standard FEBID continuum model, which primarily models the coverage of precursor molecules under electron irradiation, the new model introduces the formation of ligand fragments created during the electron-induced dissociation of precursor molecules. This extension, achieved by introducing a second surface reaction describing the fate of the ligand, expands the parameter space to describe FEBID processes mathematically.

By incorporating these additional surface species, the model allows for analysis of both the growth rate and the time dependence of deposition yield. The model has been compared with experimental data for a chromium carbonyl and a silver carboxylate precursor, showing qualitative agreement for both growth rate as a function of dwell time and metal content as a function of spatial electron flux variations. Although the model does not account for all sub-processes occurring during electron beam irradiation of adsorbed precursor molecules, it represents a significant step toward a better understanding of experimental results related to the composition of FEBID materials as a function of external parameters, particularly for single-ligand precursors. Its intentional simplicity also makes it well-suited for direct integration into 3D printing software based on the FEBID continuum model, potentially enabling reliable tailoring of both material composition and growth rates for FEBID-nanoprinted 3D nanostructures.

**Acknowledgements**

This project was supported by the European Community under the Horizon 2020 Program, Contract No. 101001290 (3DNANOMAG). J.J and I.U research was conducted with the financial support of EU Horizon 2020 Marie Curie-Sklodowska Innovative Training Network "ELENA", grant agreement No 722149, and of the Swiss National Science Foundation by the COST-SNSF project IZCOZ0_205450.
The authors thank Klaus Edinger, Christian Felix Hermanns and Reiner Becker from Carl Zeiss SMT GmbH (Rossdorf) for training and help in experiments using the prototype mask repair tool. The authors want to thank Czesław Kapusta and Aleksandra Szkudlarek from AGH University of Cracow and Luisa Berger from Empa for fruitful discussions.

**Data Availability Statement**

The data that support the findings of this study are openly available in [zenodo.org] at:

**Conflict of Interest**

The authors declare no conflicts of interest.

**ToC**

Nanoprinting using focused electron beam induced deposition (FEBID) with metalorganic molecules (ML) offers unparalleled spatial resolution and flexibility in 3D shape design. Here we address the key feature of nanoprint material - its metal composition. Including the fate of detached ligands (L) into the continuum FEBID model guides nanoprint operation and precursor design towards high metal content material.

**Ligand co-deposition in focused electron beam induced nanoprinting: a predictive composition model**

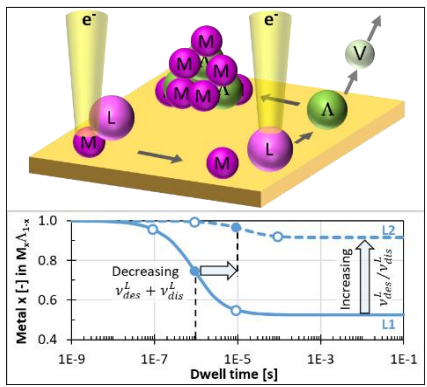